\documentclass[10pt,english,aps,prl,twocolumn,superscriptaddress]{revtex4}
\usepackage[T1]{fontenc}
\usepackage[latin9]{inputenc}
\usepackage{verbatim}
\usepackage{amssymb}
\usepackage{graphicx}
\usepackage{esint}
\usepackage{graphicx}
\usepackage{color}

\newcommand{\ket}[1]{\left| #1 \right>} 
\newcommand{\bra}[1]{\left< #1 \right|} 
\newcommand{\vsigma}{\mbox{\boldmath $\sigma$}}

\def\eea{\end{eqnarray}}

\def\bk{{\bf k}}

\def\bq{{\bf q}}

\def\9{\rangle}
\def\6{\langle}
\def\beq{\begin{equation}}
\def\eeq{\end{equation}}
\def\be{\begin{equation}}
\def\ee{\end{equation}}
\def\bea{\begin{eqnarray}}
\def\eea{\end{eqnarray}}

\def\bk{{\bf k}}

\def\bq{{\bf q}}

\makeatletter

\@ifundefined{textcolor}{}
{%
 \definecolor{BLACK}{gray}{0}
 \definecolor{WHITE}{gray}{1}
 \definecolor{RED}{rgb}{1,0,0}
 \definecolor{GREEN}{rgb}{0,1,0}
 \definecolor{BLUE}{rgb}{0,0,1}
 \definecolor{CYAN}{cmyk}{1,0,0,0}
 \definecolor{MAGENTA}{cmyk}{0,1,0,0}
 \definecolor{YELLOW}{cmyk}{0,0,1,0}
}
\def\be{\begin{equation}}
\def\ee{\end{equation}}

\makeatother

\usepackage{babel}
\begin{document}

\title{Lighting up topological insulators: large surface photocurrents from magnetic superlattices}

\author{Netanel H. Lindner}
\affiliation{Department of Physics, Technion, Haifa 32000, Israel}

\author{Aaron Farrell}
\affiliation{Department of Physics and the Centre for Physics of Materials, McGill University, Montreal, Quebec,
Canada H3A 2T8}
\affiliation{Institute of Quantum Information and Matter, Dept. of Physics, Caltech,
Pasadena, CA 91125}

\author{Eran Lustig}
\affiliation{Department of Physics, Technion, Haifa 32000, Israel}

\author{Felix von Oppen}
\affiliation{\mbox{Dahlem Center for Complex Quantum Systems and Fachbereich Physik,
Freie Universit\"at Berlin, 14195 Berlin, Germany}}

\author{Gil Refael}
\affiliation{Institute of Quantum Information and Matter, Dept. of Physics, Caltech,
Pasadena, CA 91125}

\begin{abstract}
The gapless surface states of topological insulators (TI) can potentially be
used to detect and harvest low-frequency infrared light. Nonetheless, it
was shown that significant surface photocurrents due to light with frequency
below the bulk gap are rather hard to produce. Here we demonstrate that
a periodic magnetic pattern added to the surface dramatically enhances surface
photocurrents in TI's. Moreover, the sensitivity of this set-up to the wavelength of the incident light can be optimized by tuning the geometry of the magnetic pattern. The ability to produce substantial
photocurrents on TI surfaces from mid-range and far-infrared light could
be used in photovoltaic applications, as well as for detection of
micrometer wavelength radiation. For light of wavelength greater than 15$\mu$m we estimate that at room temperature, a detector based on the effect we describe can have a specific detectivity as high as 10$^7$ cm$\sqrt{\text{Hz}}$/W (i.e. 10$^9$ Jones). The device can therefore operate at much larger wavelengths than existing infrared detectors, while maintaining a comparable figure of merit.
\end{abstract}
\maketitle

\section{Introduction}

Light-matter interactions are central to modern science and technology.
It is the principle at the heart of many solid-state material probes,
and at the same time, it is an important ingredient in our energy economy, particularly
through photovoltaic harvesting of solar energy. A challenging problem
of solar energy is how to harness the infrared (IR) part of the spectrum. This could apply to the solar radiation, as well as to Earth's radiation, which is almost exclusively in the infrared, and continuously has the same energy flux as the sun \cite{Capasso}. Similarly, electric mid and far infrared detection is essentially limited to a single type of material: HgCdTe alloys. Additional platforms are likely to be competitive in certain temperature and frequency regimes.

 Efforts to extend the spectrum accessible in photovoltaics concentrated on
new low band-gap materials; organics \cite{LowBandgapPolymer,Bundgaard,Park2009,lunt2011}, as well as carbon nanotubes \cite{Ago,Kymakis,Strano2012} were shown capable of IR harvesting, albeit with a small efficiency. Another approach
utilizes plasmonics as an intermediate step between IR and currents
in a semiconductor \cite{Atwater0,Atwater1,Atwater2,arXiv:1412.2658}.

When mentioning new materials for IR harversting, topological insulators \cite{FuKane,Zhang_3D, MooreBalents, Hseih, Xia} immediately
come to mind. On the one hand, they have a unique response to electromagnetic
fields \cite{qi2008,qi2009,Essin}. More
importantly, their mid-gap surface states exhibiting
spin-momentum locking raised hopes that
surface photocurrents could easily be produce by irradiation with  circularly-polarized light. These hopes have gone
unfulfilled. Even when including a series of modifications to the
band structure likely to appear in real materials, such as warping,
band curvature, or a uniform magnetic field, the photocurrents
produced in response to sub-bandgap light were shown to be remarkably minute,
even when a high-intensity laser is considered \cite{hosur2011,JunckRefael}.
The only scheme for producing a photo-voltage so far relied on the unique thermoelectric effects associated with a Dirac cone dispersion \cite{Jarillo-Herrero}.

In this manuscript we describe how to turn a topological insulator
surface with a simple Dirac dispersion into a photocurrent rectifier.
We show that by adding a magnetic coating with a spatially periodic
magnetic texture, the TI produces a significant surface photocurrent in response
to circularly polarized light in the IR regime. This effect should,
in principle, allow making diode-free IR sensitive photocells from
topological insulator films. We discuss application of the effect to room temperature infra-red detection, and show that it can lead to a detector operating at much larger wavelengths then those available with existing technologies.  Beyond such applications, the effect can be used to
investigate the unique properties of TI surfaces using non-ionizing
light (as in \cite{mciver2011}).

The paper is organized as follows. In Sec.~\ref{sec: summary}, we give a description of the device and summarize our main results. In Sec.~\ref{sec: model} we present the model describing the magnetically patterned TI surface. The symmetries of the model are discussed in Sec.~\ref{sec: symmetries}. In Sec.~\ref{sec: calculation} we derive the equations describing the photocurrent response of the device. We consider the implications of the symmetries on the photocurrent response (Sec.~\ref{sec: symm implications}), and find the conditions under which a large photocurrent response is obtained. A perturbative calculation of the photocurrent response is given in Sec.~\ref{sec: perturbative}.  Our main results for the frequency dependent photocurrent response of the device are given in Sec.~\ref{sec: response}. Several applications of the device, and in particular, room temperature infra-red detection are discussed in Sec.~\ref{sec: applications}. We close with concluding remarks in Sec.~\ref{sec: conclusions}.


\section{The proposed device and summary of the main results}
\label{sec: summary}

\begin{figure}
\includegraphics[width=5cm]{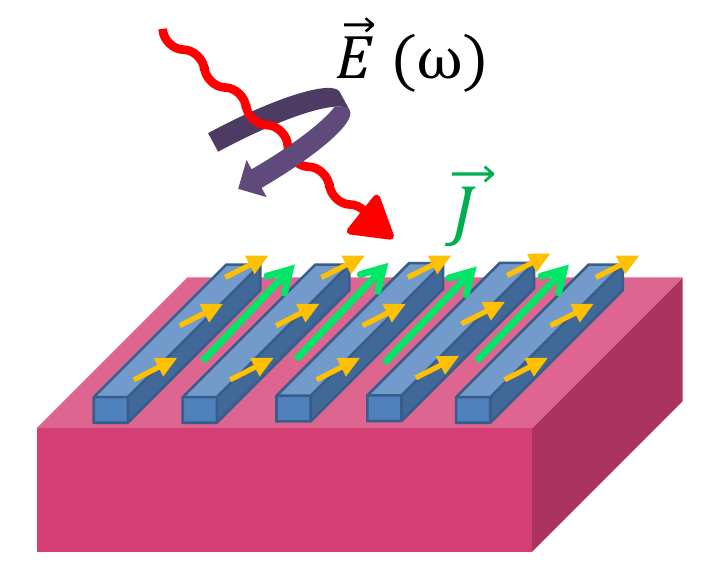}
\caption{Proposed scheme for achieving a photovoltaic effect on a topological-insulator surface, coated by a magnetic grating. When the magnetization (depicted as yellow arrows)
breaks both rotation and reflection symmetries, circularly polarized light induces a photocurrent (green) in the direction parallel to the stripes.\label{fig: scheme and bandstructure}}
\end{figure}

The device we propose and analyze in this paper consists of a bulk three dimensional topological insulator, whose surface is coated with stripes of magnetic material, see Fig.~\ref{fig: scheme and bandstructure}. We consider  magnetic stripes which are evenly spaced. The stripes' spacing defines a wave vector $\bq$ in the plane of the surface and normal to the orientation of the stripes.  Via their magnetic coupling to the electrons in the surface state of the TI, the magnetic stripes break  symmetries which suppress the photocurrents in their absence. Thereby, the magnetic stripes dramatically enhance the photocurrent response of the TI's surface. The magnetic stripes are taken to be magnetically ordered in the same direction.  As we explain in Sec.~\ref{sec: symmetries}, the direction of the stripes' magnetization needs to have non-zero components both normal to the surface as well as along the vector $\bq$. The photocurrents flow parallel to the direction of the stripes (perpendicular to $\bq$), as shown in Sec.~\ref{sec: symmetries}.

The photocurrent response of the device can be described by a dimensionless, frequency dependent response function $\eta(\omega)$. In Sec.~\ref{sec: response} we demonstrate a key feature of $\eta(\omega)$: it exhibits a strong maximum at frequency $\omega\approx1.7v_F |\bq|$, where $v_F$ is the velocity associated the Dirac cone. This result has significant implications in future applications of the proposed device: the frequency corresponding to the peak sensitivity of the device can be tuned by appropriately choosing the spacing of the magnetic stripes.  In Sec.~\ref{sec: temperature}, we analyze the performance of this set-up at finite temperature and with the chemical potential tuned away from the Dirac point. This analysis gives an ``operational'' region for the device: we show that the performance of the device is not significantly reduced for temperatures up to $\hbar v_F |\bq|$, which could translate to $300K$ in practical realizations. Similarity, we show that deviations of the chemical potential from the Dirac point do not significantly hinder the the performance, as long as they remain below  $\hbar v_F |\bq|$.

Quantitative estimates for the photocurrent response in several applications are given in Sec.~\ref{sec: applications}. We estimate that the two dimensional
photocurrent density resulting from illumination with sunlight could reach $10^{-8}\frac{A}{m}$. Illumination with a conventional laser beam can yield currents of the order of $10^{-4}\frac{A}{m}$. A particularly appealing application of the device is room temperature detection of infra-red radiation. We explore the potential of this system to detect black-body radiation emitted at a variety of different source temperatures. We conclude that the device may be able to detect black-body radiation of objects at room temperature while itself being at a comparable temperature. Finally, we explore several theoretical figures of merit for the device as a room temperature IR detector. In particular we calculate the device's external quantum efficiency and its specific detectivity, which gives its normalized signal to noise ratio  \cite{sensors3} . Near room temperature and with peak sensitivity tuned to wavelengths near $15\mu$m we estimate a quantum efficiency of 0.01$\%$ and a specific detectivity $\sim 10^{7}$ cm$\sqrt{\text{Hz}}$/W, {\em before} any device optimization takes place. Such a detectivity compares well with the detectivity of current room temperature photo-detectors \cite{detector4}, which can usually only detect up to $10\mu\mathrm{m}$ \cite{detector1, detector2, detector3, detector4, detector5, detector6}. Importantly, the proposed device has the potential to be functional for wavelengths greater than $15\mu\mathrm{m}$.  Our findings therefore support the idea that this set-up may be promising for room temperature detection of long wavelength infrared radiation.

\section{Minimal model for surface photocurrent rectification}
\label{sec: model}

Our photocurrent rectification scheme emerges from the minimal model of a surface of a 3D topological insulator (TI). With the surface lying in the $xy$ plane, the Hamiltonian describing the surface electrons is
\begin{eqnarray}
H_{0} & = & \label{eq: H0}v_{F}\left(p_x\sigma^y-p_y\sigma^x\right),
\end{eqnarray}
where $\sigma^{x},\sigma^{y}$ are
Pauli matrices, and $\mathbf{p}=(p_x,p_y)=\left(\frac{\hbar}{i}\frac{\partial}{\partial x},\,\frac{\hbar}{i}\frac{\partial}{\partial y}\right)$. This model is clearly time-reversal and rotationally invariant, ${\cal T}H_{0}{\cal T}^{-1}=U_{\phi}H_{0}U_{\phi}^{\dagger}=H_{0}$, with the symmetry operators
\begin{equation}
{\cal T}=i\sigma_{y}K,\hspace{1cm} U_{\phi}=e^{i\phi\left(\sigma^z/2+L_z/\hbar\right)}. \label{eq:rotation}
\end{equation}
Here, $K$ denotes complex conjugation, $L_z=x p_y-y p_x$ is the orbital angular momentum normal to the surface, and $\phi$ the angle of rotation.

These two symmetries immediately imply no current response to incident light at normal incidence. Time-reversal invariance requires that the incident beam is circularly polarized to see any response. Since circularly polarized light, however, has no preferred direction on the surface, the rotational symmetry rules out any net photocurrent from forming. In materials such as $\mathrm{Bi_{3}Se}_{2}$, the lattice structure
reduces the full $SO(2)$ rotational symmetry to a $C_{3}$ symmetry,
with $\phi=2\pi/3$ in Eq.\ (\ref{eq:rotation}). This allows $H_0$ to have
a trigonal warping term \cite{fu2009}. However even with the reduced symmetry, no photocurrents are possible \cite{hosur2011,JunckRefael}.

Next, we consider a magnetic grating structure deposited on the surface, see Fig.~\ref{fig: scheme and bandstructure}.  Consider strips of a ferromagnetic
material set parallel to the $y$ axis, and placed periodically with a wave number $\mathbf{q}=(q,0)$. We model the proximity-induced ferromagnetic interaction on the surface electrons by
\begin{equation}
V= \mathbf{u}\cdot{\vsigma}\cos(q x).  \label{eq: magnetic perturbatoin}
\end{equation}
Once the magnetic structure is introduced, it is convenient to enumerate the eigenstates of the full Hamiltonian, $H=H_0+V$ within the reduced Brillouin zone (BZ) in terms of the quasi-momentum $k_{x}\in[-\frac{q}{2},\frac{q}{2}]$ in x-direction, using Greek indices to denote the bands. Thus denote the eigenstates as $\ket{k_x,k_y;\alpha}$. We use the convention that conduction bands are enumerated by $\alpha>0$ and valence bands by $\alpha<0$, as illustrated in Fig.\ \ref{fig: bandstructure}

\section{Symmetry considerations of the modified surface}
\label{sec: symmetries}
The addition of the magnetic strips on the surface alters its symmetries. Time-reversal symmetry remains, as long as we consider a modified operator which concatenates time reversal with a spatial translation: $\tilde{\cal T}={\cal T}M$ with $MxM^{\dagger}=x+\pi/q$. The eigenstates and energies transform as
\begin{equation}
\ket{-k_x,-k_y;\alpha}=\tilde{\cal T}\ket{k_x,k_y;\alpha}
,E_{-k_x,-k_y;\alpha}=E_{k_x,k_y;\alpha}\label{TR}
\end{equation}
Particle-hole symmetry is also present. First define $\Pi_a$ as the spatial reflection operator about the $a=x,y$ directions, e.g., $\Pi_x x \Pi_x=-x$. Now, ${\cal C}=\Pi_x\Pi_y {\cal T}$ implements:
\be
\ket{k_x,k_y;-\alpha}={\cal C}\ket{k_x,k_y;\alpha}
,E_{k_x,k_y;\alpha}=-E_{k_x,k_y;-\alpha}.\label{PH}
\end{equation}
Additional symmetries appear restricted due to the arbitrary form of $V$.

Nonetheless, a gauge transformation allows us to cancel an arbitrary $u_y$ component of $V$, and allows additional mirror symmetries. Define the gauge transformation
\be
G=\exp{\left(i\frac{u_y}{ \hbar v_F q}\sin qx\right)} .
\ee
It is easy to verify that
\begin{eqnarray}
\tilde{H}=G H G^{-1}&=&H- u_y\sigma^y\cos qx\nonumber\\
&=&H_0+(u_x\sigma^x+u_z\sigma^z)\cos qx.
\end{eqnarray}
With $u_y$ eliminated, we can construct the mirror transformation
\be
{\cal P}_x=\Pi_y K.\label{Px1}
\ee
The only term that can possibly be affected by this compounded transformation is actually invariant,
\be
\begin{array}{c}
\Pi_y K \l[\frac{\hbar v_F}{i}\l(-\sigma^x\partial_y)]K^{-1}\Pi_y\\
=\Pi_y \l[\frac{\hbar v_F}{i}\l(\sigma^x\partial_y)]\Pi_y=\frac{\hbar v_F}{i}\l(-\sigma^x\partial_y),
\end{array}
\end{equation}
so that ${\cal P}_x\tilde{H}{\cal P}_x^{-1}=\tilde{H}$. Since complex conjugation imposes $(k_x,k_y)\rightarrow (-k_x,-k_y)$, and $\Pi_y$ reverses $-k_y$ back to $k_y$, we have:
\begin{equation}
\ket{-k_x,k_y;\alpha}={\cal P}_x\ket{k_x,k_y;\alpha}, E_{-k_x,k_y;\alpha}=E_{k_x,k_y;\alpha}.\label{Px}
\end{equation}
By compounding ${\cal P}_x$ with time reversal, $\tilde{\cal T}$, we also obtain a reflection about the x-axis: $\tilde{\cal T}{\cal P}_x:(k_x,k_y)\rightarrow (k_x,-k_y)$.
Below we will first discuss the equations describing the photocurrent response of the device, and then consider the consequences of the symmetries on the resulting photocurrent.

\section{Calculation of the photocurrent response}
\label{sec: calculation}

 Within Fermi's golden rule, we expect that the photocurrent response to a particular frequency will be quadratic in the photon field. We restrict ourselves to normally incident photons, at frequencies which allow us to approximate the vector potential as spatially uniform, $\mathbf{A}(\mathbf{x},t)={\rm Re}\mathbf{A}_{0}(\omega)e^{i\omega t}$. Then, the $k,m,n=x,y$ component of the photocurrent is quite generally given by:
\begin{equation}
j_{k}(\omega)=\frac{e\tau}{4\hbar \omega^{2}} E_{m}(\omega){\cal Q}_{kmn}(\mathbf{\omega})E_{n}^{*}(\omega). \label{eq: total current1}
\end{equation}
Here, repeated indices are summed, ${\bf E}(\omega)=-i\omega{\bf A_0}$. Also, in Eq.~(\ref{eq: total current1}) we assume that the current decays on a time scale $\tau$.
Quite remarkably, in the presence of a periodic structure of magnetic strips lying along the y-axis, we will find that there is only one independent element of ${\cal Q}_{kmn}(\mathbf{\omega})$ which is nonzero: \
\begin{displaymath}
{\cal Q}(\omega)_{y,x,y}={\cal Q}(\omega)_{y,y,x}^{*}=-{\cal Q}(\omega)_{y,y,x}.
\end{displaymath}

To calculate $ {\cal Q}_{kmn}(\mathbf{\omega})$, we first write the surface photon-electron interaction, which we assume follows the minimal coupling prescription:
\begin{equation}
\hat{H}_{int}=e\frac{\partial H_{0}}{\partial\mathbf{p}}\cdot\mathbf{A}(\mathrm{\mathbf{x}},t).
\end{equation}
{The presence of the photon field can either excite electrons to a higher sub band or allow them to relax to a lower sub band through emission. Taking these possibilities into account we have the following result for ${\cal Q}_{kmn}(\omega)$
\begin{equation}
{\cal Q}_{kmn}(\omega)=\int \frac{d^2k}{(2\pi)^2}\sum_{\alpha,\beta}{\cal Q}_{kmn}^{\alpha \beta}(\mathbf{k},\omega, T, \mu),
\end{equation}
where ${\cal Q}_{kmn}^{\alpha \beta}(\mathbf{k},\omega, T,\mu)$ describes the excitation/relaxation of electrons at momentum $\mathbf{k}$ and temperature $T$. An application of Fermi's golden rule yields:
\begin{equation}
\begin{array}{c}
{\cal Q}_{kmn}^{\alpha\beta}(\mathbf{k,\omega}, T,\mu) = {\hat\mathbf{x}_{k}}\cdot\left(\mathbf{v}_{\mathbf{k}}^{(\alpha)}-\mathbf{v_{k}^{(\beta)}}\right)M_{mn}^{\alpha\beta}(\mathbf{k})\\
\times2\pi \mathrm{\delta(E_{\mathbf{k}}^{(\alpha)}-E_{\mathbf{k}}^{(\beta)}-\omega)}(n^0_{\mathbf{k},\beta}-n^0_{\mathbf{k},\alpha}),\label{qdef2}
\end{array}
\end{equation}
where $n^0_{\mathbf{k},\beta}$ is a Fermi function at a temperature $T$ and chemical potential $\mu$ and the velocities in the band $\alpha$ are given by  $\mathbf{v}_{\mathbf{k}}^{(\alpha)}=\bra{\mathbf{k};\alpha}\frac{\partial H_{0}}{\partial\mathbf{p}}\ket{\mathbf{k};\alpha}$, and the matrix elements are given by
\be
M_{mn}^{\alpha\beta}(\mathbf{k})=\bra{\mathbf{k},\alpha}\Gamma_{m}\ket{\mathbf{k},\beta}\bra{\mathbf{k},\beta}\Gamma_{n}^{\dagger}\ket{\mathbf{k},\alpha},
\label{eq: trace formula}
\ee
with  $\Gamma_{m}=e\frac{\partial H_{0}}{\partial\mathbf{p}}\cdot\hat{\mathbf{x}}_{m}$.
From this definition it is clear that $M_{mn}^{\alpha\beta}(\mathbf{k})=\left(M_{nm}^{\alpha\beta}(\mathbf{k})\right)^*$, i.e., it is hermitian. At zero temperature and with the chemical potential tuned to the Dirac point the valance sub bands are entirely full and the conduction sub bands completely empty. In this case only electron excitation is possible and we may excite electrons from any sub band of the valence band to any sub band of the conduction band, and at any momentum. Therefore:
\begin{equation}
{\cal Q}_{kmn}(\omega)=\int \frac{d^2k}{(2\pi)^2}\sum_{\alpha>0,\beta<0}{\cal Q}_{kmn}^{\alpha \beta}(\mathbf{k},\omega),
\end{equation}
where ${\cal Q}_{kmn}^{\alpha\beta}(\mathbf{k},\omega)={\cal Q}_{kmn}^{\alpha\beta}(\mathbf{k,\omega}, T\to0, \mu=0) $ accounts for excitations from the valence band $\beta<0$ to the conduction band $\alpha>0$ at momentum $\mathbf{k}$. It is this limiting case that we will develop first, and then move on to discuss how temperature and chemical potential effect these results. Setting $T\to0$ and $\mu=0$ in Eq.\ (\ref{qdef2}) now gives
\begin{equation}
\begin{array}{c}
{\cal Q}_{kmn}^{\alpha\beta}(\mathbf{k,\omega}) = {\hat\mathbf{x}_{k}}\cdot\left(\mathbf{v}_{\mathbf{k}}^{(\alpha)}-\mathbf{v_{k}^{(\beta)}}\right)M_{mn}^{\alpha\beta}(\mathbf{k})\\
\times2\pi \mathrm{\delta(E_{\mathbf{k}}^{(\alpha)}-E_{\mathbf{k}}^{(\beta)}-\omega)},\label{qdef}
\end{array}
\end{equation}}

\subsection{Implications of the symmetries}
\label{sec: symm implications}
The calculation of the photocurrent response in the presence of the magnetic texture can now follow. Significant simplifications can be made by taking into account the symmetries discussed in Sec.~\ref{sec: symmetries}. We first define
\begin{equation}
{\cal \widetilde{Q}}_{kmn}^{\alpha\beta}(\mathbf{k})=\sum_{\sigma,\sigma'=\pm 1}{\cal Q}_{kmn}^{\alpha\beta}(\sigma k_{x},\sigma'k_{y}),
\end{equation}
which sums the contributions of the four mirror-related momenta, $\left(\pm k_{x,}\pm k_{y}\right)$, and is defined for $k_{x},k_{y}>0$. This definition takes into account all symmetry-related current cancellations.  Assuming that $u_y$ has been gauged away, we use Eqs.\ (\ref{TR}) and (\ref{Px}) to connect the contributions arising from the four momenta $(\pm k_x,\,\pm k_y)$. Due to these symmetries, along with the particle-hole transformation, Eq.\ (\ref{PH}), we have:
\begin{equation}
\begin{array}{c}
v_{x}^{\alpha}(\sigma k_{x},\sigma'k_{y})=\sigma v_{x}^{\alpha}(k_{x},k_{y}),\\
v_{y}^{\alpha}(\sigma k_{x},\sigma' k_{y})=\sigma' v_{y}^{\alpha}(k_{x},k_{y}),\\
\mathbf{v}^{\alpha}(\mathbf{k})=-\mathbf{v}^{-\alpha}(\mathbf{k})\label{vsym}
\end{array}
\end{equation}
for $\sigma,\sigma'=\pm 1$.

\begin{figure}[!t]
\includegraphics[width=7cm]{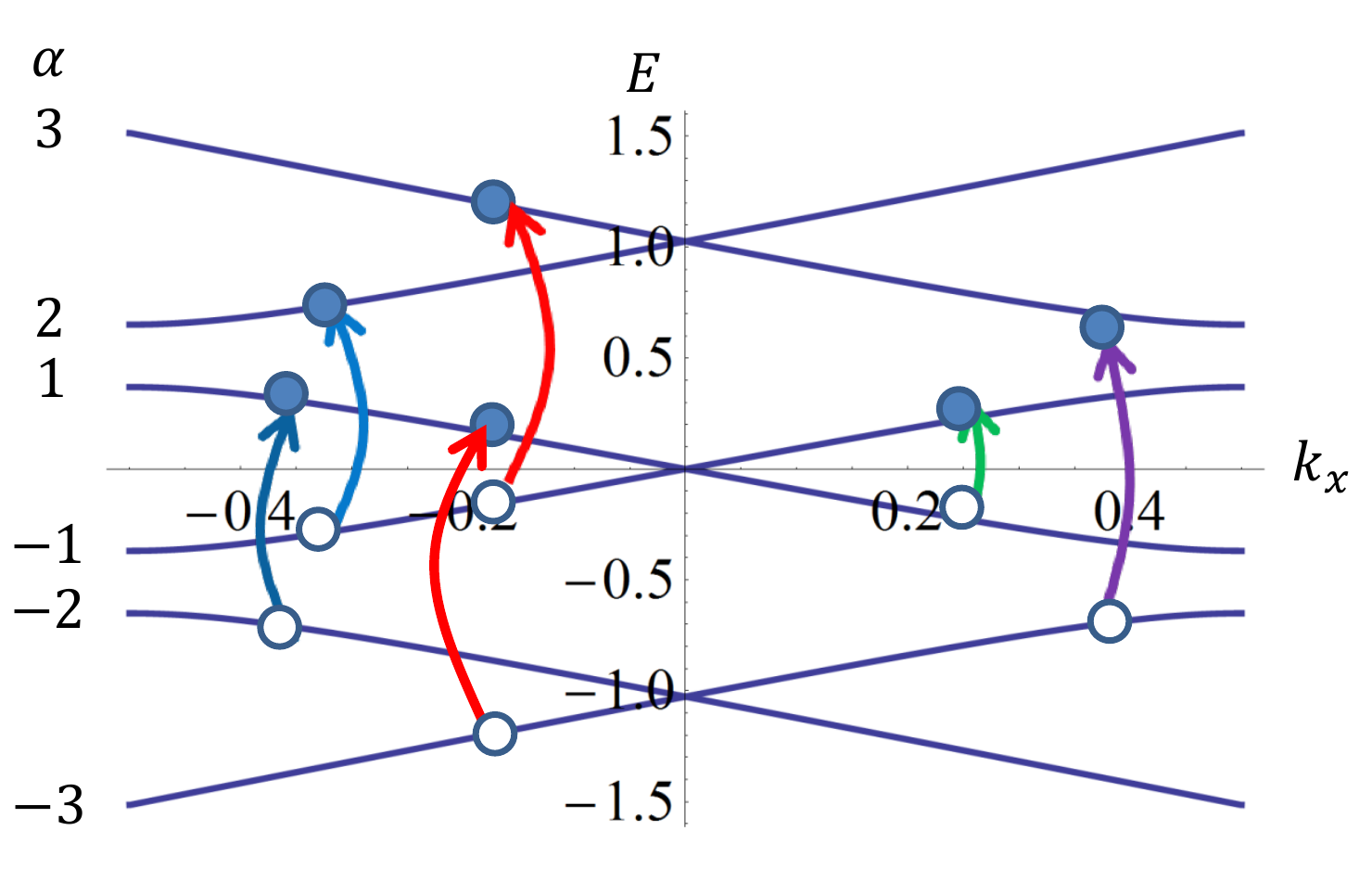}
\vspace{-0.3cm}
\caption{Effective bandstructure of the proposed heterostructure, cut along the line $k_{y}=0$ (units such that $|\mathbf{q}|=1$). Band indices are shown on the left. Transitions yielding a negative (positive) contribution to ${\cal Q}_{yxy}(\omega)$ are shown in red, blue (purple, green). The corresponding momentum dependent ${\cal Q}^{\alpha,\beta}_{yxy}(\omega)$ are given in Fig. \ref{fig: Q tensors} of the appendix.}
\label{fig: bandstructure}
\end{figure}

The same symmetries, applied to the matrix elements yield the relations
\begin{eqnarray}
\tilde{\cal T}:\,\,\,M_{mn}^{\alpha\beta}(\mathbf{-k}) & = & M_{nm}^{\alpha\beta}(\mathbf{k})\nonumber \\
{\cal P}_x:\,\,\,M_{xy}^{\alpha\beta}(-k_{x},k_{y}) & = & -M_{yx}^{\alpha\beta}(\mathbf{k})\nonumber \\
{\cal P}_x:\,\,\,M_{nn}^{\alpha\beta}(-k_{x},k_{y}) & = & M_{nn}^{\alpha\beta}(\mathbf{k}).
\label{eq: transformation M}
\end{eqnarray}
The diagonal elements, $M_{nn}^{\alpha\beta}(\mathbf{k})$, are the same at all four points $(\pm k_x,\,\pm k_y)$. This makes the contribution of these points to a current in any direction cancel identically, since the velocities obey the mirror symmetries in Eq.\ (\ref{vsym}).  From Eqs.~(\ref{vsym}) and (\ref{eq: transformation M}), we find that the only nonzero elements of the tensor ${\cal \widetilde{Q}}_{kmn}^{\alpha\beta}(\mathbf{k})$ are
\begin{eqnarray}
{\cal \widetilde{Q}}_{yxy}^{\alpha\beta}(\mathbf{k})&=&8\pi i \left[(v_y^{|\alpha|}(\mathbf{k})+ v_y^{|\beta|}(\mathbf{k})\right] \mathrm{Im}\left[M_{xy}^{\alpha\beta}(\mathbf{k})\right]\\ \nonumber &\times&\mathrm{\delta(E_{\mathbf{k}}^{(\alpha)}-E_{\mathbf{k}}^{(\beta)}-\omega)}
\label{eq: nonzero M}
\end{eqnarray}
and ${\cal \widetilde{Q}}_{yyx}^{\alpha\beta}(\mathbf{k})=-{\cal \widetilde{Q}}_{yxy}^{\alpha\beta}(\mathbf{k})$.  These conclusions confirm our claim regarding the photo-response tensor, ${\cal Q}_{kmn}(\omega)$ defined in Eq.~(\ref{qdef}): It has only one independent nonzero contribution, ${\cal Q}_{yxy}(\omega)=-{\cal Q}_{yyx}(\omega)$, which is imaginary. This implies that the current in the $x$ direction vanishes, i.e., the photocurrent is always parallel to the magnetic pattern. Furthermore, this current is only induced by the circular component of the incident light.

This result also lets us determine what magnetic patterning vector $\mathbf{u}$ is necessary for a finite response. As it turns out, having either $u_x=0$ or $u_z=0$ leads to $\mathrm{Im}M_{xy}^{\alpha\beta}=0$, and to a vanishing response. To see this, consider the composite transformation $U=\Pi_x\Pi_y\sigma^z \tilde{\cal T}$. The first part of the transformation, $\Pi_x\Pi_y\sigma^z$, implements a $\pi$ rotation on the bare model, $H_0$, and leaves it invariant. If $u_x=u_y=0$, then $H^{(z)}=H_0+u_z\sigma^z\cos(qx)$ is also invariant this transformation. $\tilde{\cal T}$ then  leaves $H^{(z)}$ invariant, and reverses momentum directions. Together, they make an anti-unitary transformation which leaves the momentum $\mathbf{p}$ invariant. Its effect on the transition matrix is $M^{\alpha\beta}_{mn}(\mathbf{k})=M^{\alpha\beta}_{nm}(\mathbf{k})=M^{\alpha\beta}_{mn}(\mathbf{k})^*$. The same relation is obtained also for the case $u_z=0$ with a finite $u_x$, with $UM$ used instead of $U$ (with M the half-period translation operator). Thus both $u_z$ and $u_x$ must be finite for a finite photo-response.

\subsection{Perturbative photocurrent calculation}
\label{sec: perturbative}

The summed momentum-specific photocurrent contributions, ${\cal \widetilde{Q}}_{kmn}^{\alpha\beta}(\mathbf{k})$, can be found analytically to lowest order in the strength of the magnetic texture. To do so, we expand the eigenstates of the Hamiltonian that appear in the definition of $M_{mn}^{\alpha\beta}(\mathbf{k})$ in Eq.\ (\ref{eq: trace formula}), and also separate the current inducing processes ${\cal \widetilde{Q}}_{kmn}^{\alpha\beta}(\mathbf{k})$ according to channels of interband scattering. In terms of momenta in the \textit{extended} BZ, the possible scattering processes to order $V^2$ are $\bk\to\bk+q$, $\bk\to\bk-q$, and $\bk\to\bk$.  The resulting photocurrent can be written as
\begin{equation}
{\cal \widetilde{Q}}^{\rm ext}_{yxy}(\mathbf{k})  =  2\pi u_x u_z(e v_F)^2 \sum_{\lambda=0,+,-} F_\lambda(\mathbf{k})\delta_\lambda(\omega,\mathbf{k}) ,
\label{eq: q tensors 2nd order}
\end{equation}
where the functions $F_\lambda$, $\lambda=+,-,0$  account for the above scattering processes and are given by
\begin{eqnarray}
F_0(\mathbf{k}) & = & v_{F}\frac{-512ik_{y}^{2}k_{x}^{2}}{|\mathbf{k}|^{2}(-4k_{x}^{2}+\mathbf{q}^{2})^{2}}\nonumber \\
F_{\pm}(\mathbf{k}) & = & v_{F}\frac{8ik_{y}^{2}q^{2}}{|\mathbf{k}|^2\,|\mathbf{k\pm q}|^2(|\mathbf{k \pm q}|-|\mathbf{k}|){}^{2}}.\label{eq: result 2nd order}
\end{eqnarray}
The delta functions in Eq.~(\ref{eq: q tensors 2nd order}) were abbreviated to $\delta_\lambda(\omega,\mathbf{k})=\delta(E_{\mathbf{k}+\lambda \bq}+E_{\mathbf{k}}-\omega)$. The momentum integrated response tensor becomes ${\cal Q}_{kmn}(\omega)=\int \frac{d^2k}{2\pi} {\cal \widetilde{Q}}^{\rm ext}_{kmn}(\mathbf{k})$, where the integral is taken over the $k_x,k_y>0$ quadrant of the \textit{extended} BZ \footnote{The mapping between the band index $\alpha$ and momentum $\bk$ in the reduced BZ, and the  extended zone momenta is given by $\bk;\alpha\to\bk-(-1)^{\alpha}\lfloor\frac{\alpha}{2}\rfloor \bq$. Note that the divergences cancel between $F_0(\bk)$ and $F_\pm(\bk)$}  (see the appendix for more information).

\section{Results: Photocurrent response of the proposed device}
\label{sec: response}
Our results are best expressed in terms of the intensity, $I$, of the light field. For a coherent monochromatic circularly polarized wave with electric-field amplitude $E_0$, we have $I=\epsilon_0 c E_0^2$. This yields the current response:
\begin{equation}
j_y(\omega)=\frac{e^3 v_F^2 q\tau}{\epsilon_0 c \hbar^2}\frac{I}{\omega^2}\eta(\omega)\label{eq: total current density}
\end{equation}
in terms of the dimensionless frequency-dependent response, $\eta(\omega)$, defined by ${\cal Q}_{yxy}(\omega)=2\frac{e^2 v_F^2 q}{\hbar}\eta(\omega)$. For a continuous spectrum with intensity per unit angular frequency, containing both circular polarizations, we write $I(\omega)d\omega = 2\epsilon_0 c |{\bf E}(\omega)|^2$. The total current response is then:
\begin{equation}
j_y=\frac{e^3 v_F^2 q\tau}{2\epsilon_0 c \hbar^2}\int\limits_0^{\Omega}\frac{I(\omega)}{\omega^2}\eta(\omega)d\omega\label{eq: total current density-cont},
\end{equation}
where $\Omega$ is the high-frequency cutoff.

\begin{figure}
\includegraphics[width=7.5cm]{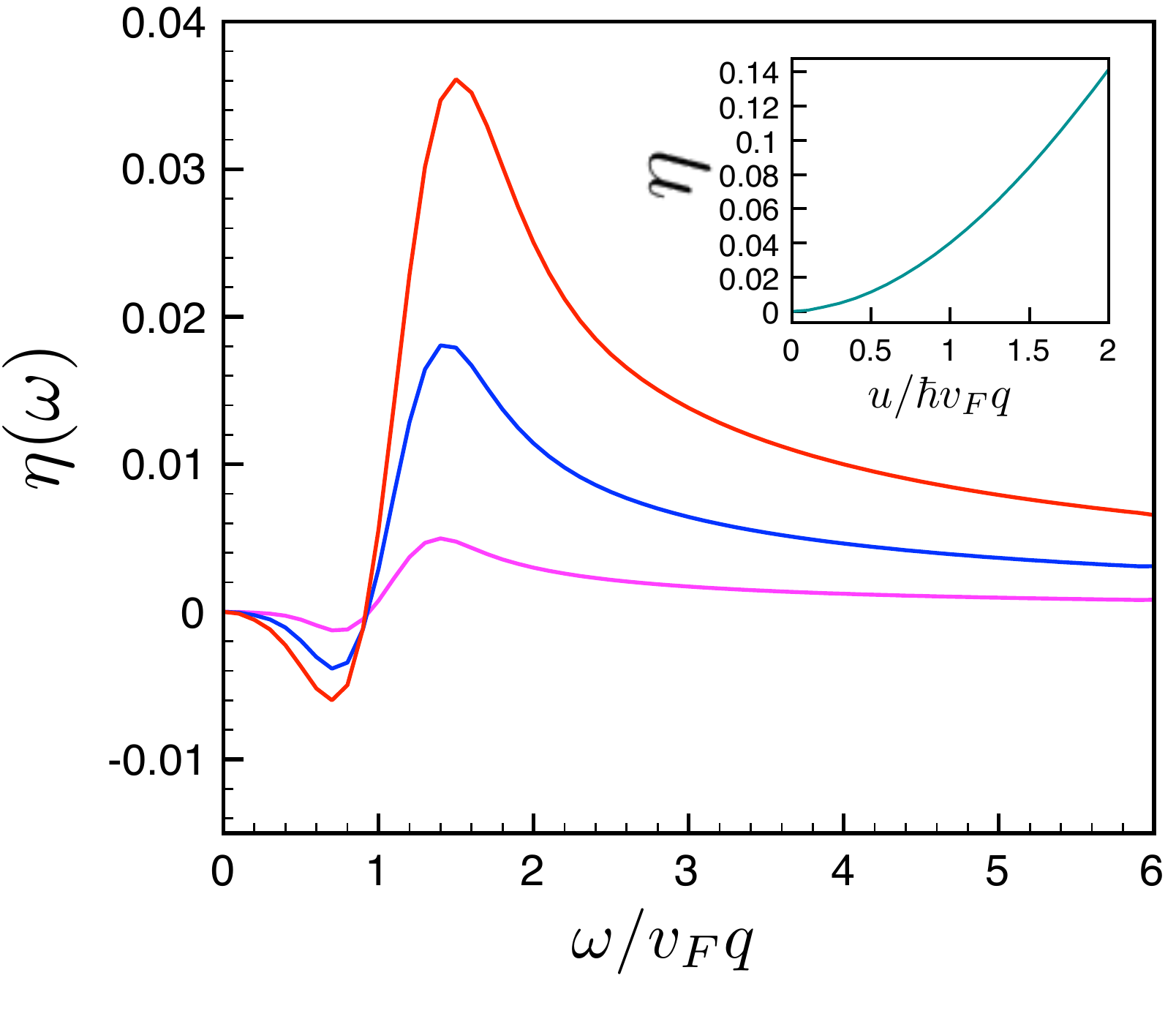}
\vspace{-0.5cm}
\caption{The dimensionless response function $\eta(\omega)$, for $u_{x}/\hbar v_{F}q=u_{z}/\hbar v_{F}q=0.1,0.2,0.3$ (purple, blue and red, respectively). The horizontal axis gives the frequency $\omega$ in units of $v_{F}q$. The inset shows the saturation value $\eta$ of the response function at high frequencies $\omega\gg v_F q$, as a function of $u/\hbar v_F q$, with $u=u_x=u_z$. }
\label{fig:The-dimensionless-response}
\end{figure}

\begin{figure*}
\includegraphics[width=17.5cm]{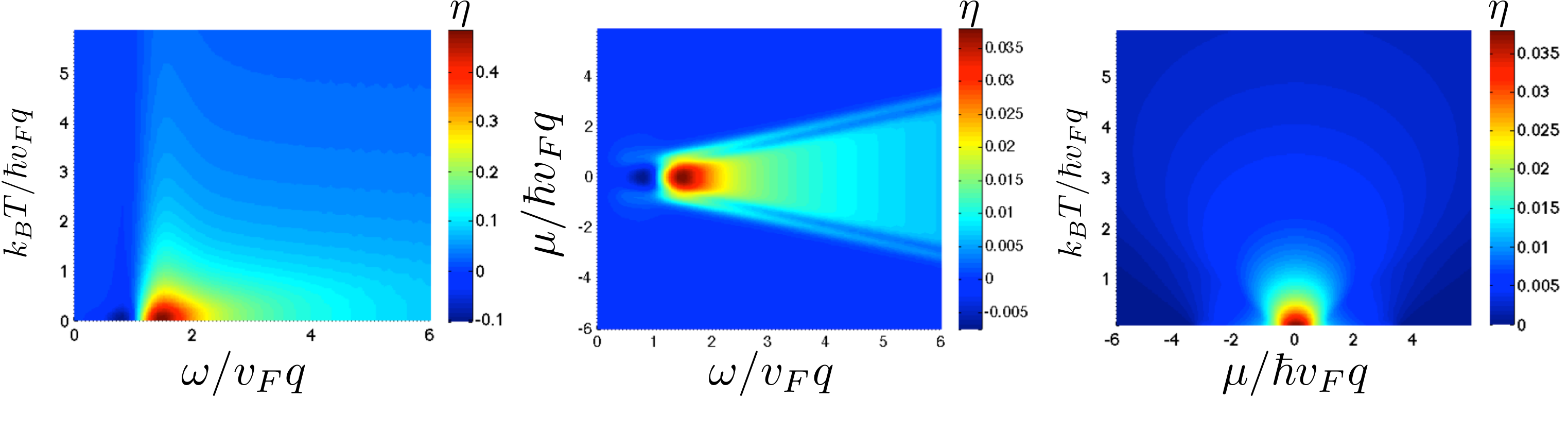}
\vspace{-0.5cm}
\caption{Device performance for various temperatures and chemical potential strengths. Left: $\eta$ plotted over a space of $T$ and $\omega$ values with $\mu=0$, middle: $\eta$ plotted for a small but finite temperature of $T=10^{-3}\hbar v_F q/k_B$ over a space of $\omega$ and $\mu$ values, and right: the peak value of $\eta$ for various different chemical potential strengths and at different temperatures. In all plots we have fixed the strength of the magnetic coupling such that $u_x=u_z=0.3\hbar v_Fq$. Notice that $\eta$ remains large up until $k_BT\sim \hbar v_F q$ indicating that the proposed device may still be functional at large temperatures. Similarly, $\eta$ is nonzero for a wide range of chemical potentials  showcasing the freedom this device has in where its Fermi level is set.}\label{fig:tempmu}
\end{figure*}

Fig.\ \ref{fig:The-dimensionless-response} displays our numerical results for the dimensionless response $\eta(\omega)$ for three magnetic patterning strengths. We make three observations: (1) Most of the contribution to the current density arises from frequencies $\omega>v_{F}q$. (2) For $\omega\gg v_F q$, the dimensionless response
$\eta(\omega)$ approaches a constant. (3) $\eta(\omega)$ changes sign at $\Omega^{*}\sim v_{F}q$. The latter observation can also be deduced from Eq.~(\ref{eq: result 2nd order}), as $F_0(\mathbf{k})$ differs in sign from the two other contributions, and dominates below $\Omega^{*}$.

Further intuition for the origin of the photocurrent distributions can be gained by studying the momentum-specific response $\tilde{{\cal Q}}_{yxy}^{\alpha\beta}(\mathbf{k},\omega)$. These are plotted in Fig.~\ref{fig: Q tensors} of the appendix, which demonstrates that the momenta contributing to the photocurrent are uniformly distributed in the reduced BZ. Furthermore, in agreement with the perturbative results,  the sign change of $\eta(\omega)$ is shown to arise due to processes involving scattering by momentum $\pm\bq$ (indicated in red and blue in Fig.~\ref{fig: bandstructure}) which dominate for $\omega>\Omega^*$; they contribute with opposite sign to momentum conserving processes (green and purple in Fig.~\ref{fig: bandstructure}), which dominate at low frequencies.

\subsection{ {Temperature and chemical potential dependence of the photocurrent response}}
\label{sec: temperature}
Next let us illuminate the potential operating regime of such a device. More specifically, let us address the question of functionality of the above device over a range of temperatures and chemical potential. Towards this end, we have evaluated the dimensionless response, $\eta(\omega)$, at several operating temperatures and with the chemical potential tuned away from the Dirac point. Our results of this calculation are summarized in Fig. \ref{fig:tempmu}.

 {By studying  Fig.~\ref{fig:tempmu} we learn several important factors for the operation of the device. First, by examining Fig.~\ref{fig:tempmu}a  we see that for temperatures $k_BT\le \hbar v_F q$  the features of $\eta(\omega)$ are not significantly changed; rather, increasing the temperature in this range seems only to moderately suppress $\eta(\omega)$. Consequently, the peak of $\eta(\omega)$ is suppressed by about $60\%$ at $k_BT=\hbar v_Fq$. For $k_BT>\hbar v_F q$ the peak becomes very flattened out and is lost. }

 {Second, in Fig.~\ref{fig:tempmu} we study the effect of the chemical potential. We see that for $\mu<\hbar v_Fq$ ($\mu=0$ signifies the Dirac point), tuning the chemical potential away from the Dirac point leads to an overall moderate suppression of $\eta(\omega)$ but has little effect on its functional form. Furthermore, we see that the value of $\eta(\omega)$ becomes almost entirely ``turned off" after a critical value of $\mu\simeq\pm \omega/2$.}

 A heuristic understanding for this behaviour is as follows. At zero temperature all states with energy above $\mu$ are empty, and those with energy below $\mu$ are occupied. Moreover, as discussed previously, the system is particle-hole symmetric. A major contribution to $\eta$ comes from electrons at energy $-\omega/2$ being excited into states at energy $\omega/2$. As $\mu$ is tuned away from zero this is changed very little until it reaches $\omega/2$ (or $-\omega/2$). At this point the transition from $-\omega/2$ to $\omega/2$ is no longer possible because the state at energy $\omega/2$ (-$\omega/2$) is full (empty). Thus the value of $\eta$ is largely suppressed after this point.

 Finally, Fig. \ref{fig:tempmu}c shows the behaviour of the peak $\eta$ value as a function of $\mu$ and $T$. This plot summarizes our main conclusion of this part of the paper: the zero temperature, zero chemical potential results are not significantly changed provided $k_BT$ and $\mu$ are within $\hbar v_Fq$ of zero. For $T$ and $\mu$ outside of this region the current is highly suppressed. This gives the appropriate operational region for such a device. In a typical TI one can expect $v_F\sim 10^5$m/s. With this and a reasonable grating pitch of $q\sim 10^8$m$^{-1}$ the operational temperature scale is set at approximately $380K$.

In the above discussion we treated the effects of finite temperature by considering its effect on the electronic occupation of the surface states. Another effect will come from phonon scattering at finite temperature. Scattering from phonons will lift the momentum conservation conditions assumed above. The strength of electron-phonon interactions on the surface of a TI is presently an active area of research \cite{relaxtime1, relaxtime2,phonons1,phonons2,phonons3,phonons4,phonons5,Parente2013}. {The role that phonons will play in this device is also an open issue. Intuitively one may expect that the phonons will scatter photoexcited electrons thereby reducing the photocurrent. It may, however, be possible to see an analogue of the phonon assisted transitions recently predicted in graphene \cite{kadi}. We leave a rigorous consideration of these two situations to future work.}  In our present treatment all scattering effects are incorporated into the relaxation time $\tau$. At low temperatures phonon modes are frozen out and scattering should be dominated by disorder \footnote{Due to the Dirac dispersion of TI surface electrons, Coulomb interactions are unable to relax a current when the chemical potential is near the neutrality point, see ref.\ \cite{Junckamp}}. As working estimates we take $\tau\sim 1$ps at low temperatures and $\tau\sim 0.1ps$ near room temperature \cite{relaxtime1,relaxtime2}.

\section{ {Applications}}
\label{sec: applications}
Here we outline several appealing practical aspects of this device. We begin by calculating some representative photocurrents for the illumination of the device by particular radiation sources. We move on to discuss the ``tunability"  of the device with $q$ and consider its application to room temperature black body detection. We close with a theoretical treatment of the figures of merit for the device as a room temperature IR detector. We show that at room temperature, the normalized signal to noise ratio (specific detectivity) is comparable with present technologies. Importantly, the device can achieve this signal to noise ratio for wavelengths which go beyond those accessible for current technologies.

\subsection{ {Solar energy}}
\label{subsec: solar}
 An  appealing application of the magnetically patterned surface is solar energy harvesting, particularly in the IR range. The intensity spectrum of the sun, for low frequencies, is approximately
given by the Rayleigh-Jeans law, $I=\frac{k_BT_{sun}}{4\pi^2c^2}\omega^2$. At the Earth's distance from the sun, at normal incidence we expect this to be suppressed by $\left(R_{sun}/R_{Sun-Earth}\right)^2\approx 2\cdot 10^{-5}$. Combined, this yields the 2d closed-circuit current expected for normally incident sunlight:
\begin{equation}
j_{y}^{(solar)}\approx \frac{e^3 v_F^2 q\tau}{2\epsilon_0 c \hbar^3} \frac{k_BT_{sun}}{4\pi^2c^2}\left(\frac{R_{sun}}{R_{Sun-Earth}}\right)^2 E_{gap}\eta_\infty,
\end{equation}
where $E_{gap}$ is the bandgap of the topological insulator hosting the Dirac cone, and $\eta_\infty$ is the constant characterizing $\eta(\omega)$ at frequencies $\omega\gg v_F q$.  {We use a scattering timescale of $\tau=1ps$, a typical bandgap of $E_{gap}\approx 0.3 eV$, and a wavenumber for the magnetic structure $q=10^{8}m^{-1}$. For low magnetic coupling, we obtain $\eta \approx 0.0345 (\frac{u}{\hbar v_{F}q})^{2}$,
see Fig.\ \ref{fig:The-dimensionless-response}. Taking a typical Fermi velocity of $v_{F}=5\cdot 10^{5}\frac{m}{sec}$,  we use $\eta_\infty=0.01$ which corresponds to a magnetic coupling of about $17 $meV. The above parameters yield $j_y^{(solar)}\approx 4\eta\times 10^{-7} A/m$ }

\subsection{Laser induced photocurrents}

The effect can also easily be explored using monochromatic laser light. Using the same parameters as above, Eq.~(\ref{eq: total current   density}) yields:
\be
j_y\approx 2\cdot 10^{21}\frac{I}{\omega^2}\eta(\omega)\;\;\frac{A m}{J sec}
\ee
For laser light of intensity $I=10^5W/m^2$ \cite{mciver2011} at angular frequency $\omega=3\cdot 10^{14}s^{-1}$, with $\eta(\omega)\sim .1$ this yields $j_y\sim 2\cdot 10^{-4}A/m$.

\subsection{ Room temperature detection of infrared radiation}
A particularly appealing application of the device is detection of infrared radiation. We now look at the question of optimal detection of thermal radiation for different emitter and device temperature. Our results show that the device can serve as an efficient room temperature detector of IR radiation. For the purposes of this discussion we will assume the radiation comes from a black body in equilibrium with its environment at a temperature $T_{BB}$. Such an object radiates at intensity $I(\omega, T_{BB})= \frac{1}{4\pi^2} \frac{\hbar \omega^3}{c^2} \frac{1}{e^{\hbar\omega/k_BT_{BB}}-1}$, which has a maximum at frequency $\omega_{peak} =  b k_B T_{BB}/\hbar$ and  $b=2.8$.

Keeping this fact in mind, we now point out the following desirable quality of the our proposed set-up: the frequency that the device is most sensitive to can be tuned by changing the grating pitch $q$, since the peak in $\eta(\omega)$ occurs at $\omega\simeq 1.7v_F q$, see Fig.~\ref{fig:The-dimensionless-response}. Note that this observation is very insensitive to temperature. Given this we now imagine fabricating our device such that the peak in $\eta(\omega)$ and the peak in the black-body spectrum coincide, this requires that we set $q=\omega_{peak}/(1.7 \hbar v_F)$.


\begin{figure}
\includegraphics[width=7.5cm]{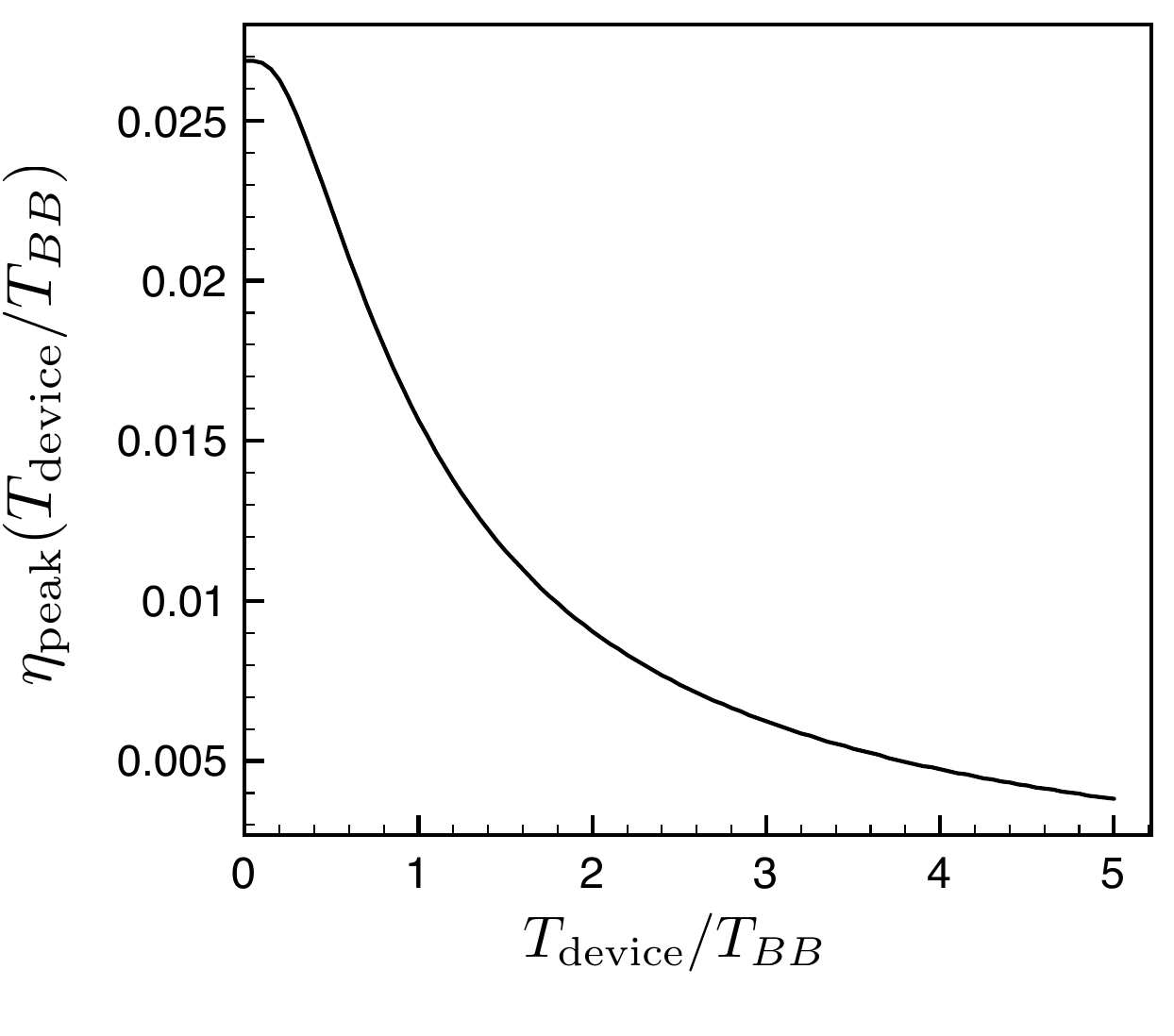}
\vspace{-0.5cm}
\caption{$\eta_{peak}$ as a function of the ratio of device temperature to black body temperature ($T_{device}/T_{BB}$). Here we have also fixed $u_x=u_z=0.3(\hbar v_F q)$. The figure shows that up to $55$\% of $\eta_{peak}$ remains intact when $T_{device}=T_{BB}$. This implies that the proposed device may be able to detect radiation from a black body at temperature $T_{BB}$ while itself being at this temperature.}
\label{fig:eta_peak_BB}
\end{figure}

 {We now gauge the performance of the system for differences in temperature between the device and the radiation source. We define $\eta_{peak}(T_{device}/T_{BB})$  as $\eta(\omega_{peak})$ when the device is set with $q=\omega_{peak}/(1.7v_F)$  and is operated at a temperature $T_{device}$ and chemical potential $\mu=0$.  We plot $\eta_{peak}(T_{device}/T_{BB})$ in Fig. \ref{fig:eta_peak_BB}. As expected, the function decreases with $T_{device}/T_{BB}$. Interestingly, we see that for $T_{device}/T_{BB}\simeq 1$, i.e., a device and black body at similar temperatures, nearly $55\%$ of the peak value of $\eta(\omega_{peak})$ remains. This is of particular interest to room temperature detection of IR radiation, where both the device and the black body are near the same temperature and where the black-body radiation lies within the gap of the TI.}

We now move on to discus the figures of merit \cite{sensors1,sensors2,sensors3} for the detector we have described above. The first is the external quantum efficiency. This figure of merit quantifies the optical absorption of the device and is defined as $E_{Q} = \frac{\hbar \omega}{e} R_{I}$ where the responsivity, $R_I$, is given by $R_I=i_{photo}/(A I_{incident})$, where $I_{incident}$ is the intensity of the incident radiation and $A=L_xL_y$ is the area of the device absorbing this radiation. In our device $L_y$ is the length parallel to the stripes and $L_x$ the length perpendicular to them.

By defining the dimensionless frequency $\omega = v_F q \bar{\omega}$ we can write
\beq\label{EQdef}
{E_{Q} =  \frac{e^2}{2\hbar c \epsilon_0} \frac{ v_F \tau}{L_y}  \frac{\eta(\bar{\omega})}{\bar{\omega}}  }
\eeq
There are several interesting pieces of information in this expression. First, we see that reducting $L_y$ leads to a higher quantum efficiency. Second, similar to the discussion above, the frequency at which the device has the highest quantum efficiency is completely tunable with the grating wave vector $q$. In other words, this frequency scale {\em is not} set by a band gap as it is in traditional semiconductor based detectors. Third, the above is a result for a single device. We could in principle layer thin films of this device in order to multiply the efficiency; the incident light not absorbed by one layer has the potential to be absorbed by other layers. Finally, at room temperate an optimized value of  $E_Q\sim 0.01\%$ is obtained using $\tau=0.1$ps and $L_y=100nm$.  {This value is independent of the wavelength $\lambda$ of the incident radiation, if the device's grating pitch $q$ is set to optimize $E_Q$}. As a comparison, the room temperature detector proposed in Ref. \cite{detector4}, functional near $\lambda\simeq 10.6\mu$m, has a quantum efficiency $\sim0.01\%$ as well.

\begin{table*}[th!]
\centering
    \begin{tabular}{ | l | l | l | p{5cm} |}
    \hline
    {\bf Device} & {\bf Temperature} & ${\bf D}^*$  & {\bf Incident $\lambda$} \\ \hline
   Magnetically Coated TI ($q=3\times10^8$m$^{-1}$)  & 300K & $ 5\times 10^6$cm$\sqrt{\text{Hz}}$/W & 10$\mu$m\\ \hline
   Magnetically Coated TI ($q=2\times10^8$m$^{-1}$)  & 300K & $ 7\times 10^6$cm$\sqrt{\text{Hz}}$/W & 15$\mu$m\\ \hline
   Magnetically Coated TI ($q=0.5\times10^8$m$^{-1}$)  & 300K & $ 1.4\times 10^7$cm$\sqrt{\text{Hz}}$/W & 31$\mu$m\\ \hline
    Graphene geometric diode rectenna\cite{detector4} & 300K & $10^{6}-10^{8}$cm$\sqrt{\text{Hz}}$/W &10.6$\mu$m  \\ \hline
   HgCdTe Photon Detector\cite{detector5, detector4} & 300K & $10^{8}$cm$\sqrt{\text{Hz}}$/W &10.6$\mu$m  \\ \hline
   Ni-NiO-Ni travelling wave MIM rectenna\cite{detector7, detector4} & 300K & $10^{9}$cm$\sqrt{\text{Hz}}$/W &10.6$\mu$m  \\ \hline
    \end{tabular}
\caption{Specific detectivities  of the device proposed in this paper compared to other devices. For our calculations we have used the estimates $L_x=1mm$, $L_y=100nm$, $I_{incident}\sim 10^{4}$W/m$^2$ and  $R_D\sim10^5\Omega$ \cite{TIresistance}}\label{table:Dstar}
\end{table*}

{The second figure of merit we wish to consider is the specific detectivity of the device. One issue with a photodetector is differentiating a photo-induced current from other ``dark" currents, i.e. those created by noise. Here we will call these noise currents $i_{noise}$.  In principle one would like the signal-to-noise ratio $i_{photo}/i_{noise}$ to be large. In practice, it is useful to define something called the specific detectivity, $D^*=\sqrt{A\Delta f} {R_I}/{i_{noise}}$ \footnote{An alternative, but equivalent, definition of the specific detectivity is the reciprocal of a measure called the noise-equivalent power normalized by the square root of the sensor's area and frequency bandwidth, in symbols $D^*=\sqrt{A\Delta f}/\text{NEP}$ with $\text{NEP}$ the noise-equivalent power. The noise-equivalent power is the incident power required to have a signal-to-noise ratio of 1}, where $A$ and $R_I$ are the area and responsivity that we defined previously and $\Delta f$ is the range of operational frequencies of the device used to measure current (used here for illustration only, as it will ultimately cancel out). We will assume our system is prone to shot and thermal noise which gives rise to\cite{sensors2} $i_{noise}=\sqrt{\left(2ei_{induced}+\frac{4k_BT}{R_D}\right)\Delta f}$. Here, $R_D$ is the resistance of the device, and $i_{induced}$ is the current from sources other than noise (e.g. photocurrent and/or the current from a bias etc.). With this model, and assuming the only induced current in the device is the photocurrent, we find the specific detectivity of our proposed device is
\begin{eqnarray}\label{detectivity}
D^* &=&  \frac{e^2}{2\hbar c \epsilon_0}  \frac{\tau}{\hbar  q}   \sqrt{\frac{L_x}{L_y}} \frac{\eta(\bar{\omega})}{\bar{\omega}^2} \\ \nonumber &\times&      \left[\frac{4k_BT}{e^2 R_D} +  \frac{e^2}{\hbar c \epsilon_0}  \frac{\tau }{\hbar  q }  \frac{\eta(\bar{\omega})}{\bar{\omega}^2} L_xI_{incident}\right]^{-1/2}
\end{eqnarray}
 where $I_{incident}$ is the incident intensity of radiation, and we have assumed a monochromatic source of light.

  {Similar to the external quantum efficiency, the detectivity of this device only depends on the frequency of incident radiation through $\eta(\bar{\omega})$ and as such can optimized by choosing $q$}. Second, near room temperature the first term on the second line of Eq. (\ref{detectivity}) dominates and} we see $D^*$ scales with $\sqrt{L_x/L_y}$ and so having a ``rectangular" device which is large in the $x$ direction compared to the $y$ direction is most beneficial. Third, we again note that the above expression is for a single device. One could imagine engineering a layered geometry of many of these devices. The signal current would increase with the number of layers whereas the noise would scale as the square root of these layers. Thus overall $D^*$ should scale like the square root of the number of layers.

 {The utility of $D^*$ is that it enables a comparison of performance across different detector technologies. We present such a comparison in Table~\ref{table:Dstar}, where we give results for the detectivity of our proposed device (for several different values of $q$) alongside $D^*$ for several other high temperature IR detectors. The table demonstrates that the specific detectivity of the proposed device compares well with recent measurements in other technologies capable of detecting IR radiation at room temperature\cite{detector1, detector2, detector3, detector4, detector5, detector6, detector7}. Importantly, the proposed device achieves these values at large wavelengths, which are beyond reach for these technologies. In fact, note that for the proposed device, $D^*$ grows with increasing $\lambda$. This is a very desirable property for building a room temperature mid and far-IR detector\cite{sensors1}.  Finally, we remind the reader that the above numerical estimates do not include any of the possible device optimization routes outlined above.}


\section{Conclusions}
\label{sec: conclusions}
The unique properties of the surfaces of topological insulators beg to be translated into practical applications. The lack of a generic photocurrent response on such surfaces so far has stifled the possibility of applications in light detection and photovoltaics. In this manuscript we demonstrated how surface magnetic patterning employs the spin-orbit locking, and allows for a substantial photocurrent response even to low-intensity sources such as the low-energy solar spectrum. The surface is naturally sensitive to photon energies below the bandgap of $0.3eV$, as opposed to semiconductor based photovoltaics, which require energies that exceed the material's bandgap. As such, this effect can be used for detection of micrometer wavelength radiation - a range with limited electric detection schemes. Our estimates give a  specific detectivity of $\sim10^{7}$ cm$\sqrt{\text{Hz}}$/W at $15\mu$m and room temperature, with the ability to go to higher wavelengths by adjusting the separation between magnetic strips. This value of the specific detectivity has room for further optimization by, e.g. creating a layered device. Present technologies with comparable room temperature detectivities are confined to wavelengths $\le10\mu$m and therefore the proposed device represents a significant potential advancement in mid-IR and far-IR detection. We note that magnetic coating of topological insulators has been experimentally demonstrated in refs.~\cite{WrayHasan,Schlenk2013,Wei2013}, as well as studied numerically using first principle calculations \cite{Zhao2014}. The use of magnetic insulators, such as the ones used in \cite{Wei2013}, will be advantageous in order to minimize effects such as absorption by the magnetic coating and electron doping of the TI surface.

Many aspects remain unexplored. To understand how the TI surface could be harnessed for solar energy harvesting, we need to understand what the natural open-circuit voltage is. In addition, we have only provided a crude account of disorder and phonon scattering effects on the surface, and completely ignored the possibility of bulk contributions at high frequencies. Lastly, we are confident that the magnitude of the effect could be improved by optimizing our device by using other magnetic patterns, or different materials. For instance, we expect that a similar affect will exist in arrays of 2d topological insulator strips, e.g., HgTe/CdTe heterostructures, put in an in-plane spatially varying field. We intend to explore at least some of these issues in future work.

\section{Acknowledgments}

We acknowledge financial support from NSF through DMR-1410435, the Packard Foundation, the IQIM - an NSF center funded in part by the Gordon and Betty Moore Foundation - and especially DARPA through FENA (Caltech), through SPP 1666 of the Deutsche Forschungsgemeinschaft and the Helmholtz Virtual Institute ``New States of Matter and Their Excitations'' (Berlin). NL acknowledges support from the CIG Marie Curie grant, the Bi-National Science Foundation and I-Core: the Israeli Excellence Center "Circle of Light" (Technion). AF acknowledges support from the Natural Sciences and Engineering Research Council of Canada through the Vanier Canada Graduate Scholarships program and McGill HPC supercomputing resources. Finally, GR and FvO acknowledge the hospitality of the Aspen Center for Physics where a portion of this work was completed.

\bibliographystyle{apsrev4-1}
\bibliography{photocurrents_refs_new_resub}

\appendix
\begin{widetext}
\section{\MakeUppercase{Photocurrent due to the solar spectrum}}

We begin with the photocurrent induced by the incident radiation

\begin{equation}
j=e\int\frac{d^{2}k}{(2\pi)^{2}}\sum_{\alpha}\left[\mathbf{v}_{\alpha,\mathbf{k}}(n_{\mathbf{k},\alpha}-n_{\mathbf{k},\alpha}^{0})\right]
\end{equation}
where $n_{\mathbf{k},\alpha}^{0}$is the equilibrium distribution
function, and $n_{\mathbf{k},\alpha}$is the distribution function
induced by the incident light. Here we derive the result for the chemical potential is
at the Dirac point and zero temperature. The generalization to finite chemical potential and temperatures will be discussed later. We model the relaxation of the system within the relaxation time approximation. Working at $\mu=k_BT=0$ all negative energy states are occupied and all positive energy states are vacant in the absence of the light. Owing to this the light must excite a negative energy electron into a positive energy state. This ultimately leads to the results

\[
(n_{\mathbf{k},\alpha}-n_{\mathbf{k},\alpha}^{0})=\tau\sum_{\beta<0}\Gamma(\mathbf{k},\beta\to\mathbf{k},\alpha)(n_{\mathbf{k},\beta}^{0}-n_{\mathbf{k},\alpha}^{0}),\qquad\alpha>0
\]
and
\[
(n_{\mathbf{k},-\beta}-n_{\mathbf{k},-\beta}^{0})=-(n_{\mathbf{k},\beta}-n_{\mathbf{k},\beta}^{0}),\qquad\beta>0.
\]
where $\Gamma(\mathbf{k},\beta\to\mathbf{k},\alpha)$ is the transition rate for an electron to move from state $(\mathbf{k},\beta)$ to state $(\mathbf{k},\alpha)$. The above considerations give us
\begin{equation}
j=e\tau\int\frac{d^{2}k}{(2\pi)^{2}}\sum_{\alpha>0,\beta<0}\left[\mathbf{v}_{\alpha,\mathbf{k}}\Gamma(\mathbf{k},\beta\to\mathbf{k},\alpha)-\mathbf{v_{\beta,\mathbf{k}}}\Gamma(\mathbf{k},\beta\to\mathbf{k},\alpha)\right].
\end{equation}

As an approximation of the transition rates we use Fermi's golden rule which gives
\begin{equation}
\Gamma(\mathbf{k},\beta\to\mathbf{k},\alpha)=\frac{|\langle\mathbf{k},\alpha|H_{int}(\omega)|\mathbf{k},\beta\rangle|^{2}}{\hbar}2\pi\delta(E_{\mathbf{k},\alpha}-E_{\mathbf{k},\beta}-\omega)
\end{equation}
for a time dependent Hamiltonian containing a single frequency.

The interaction hamiltonian is written as

\begin{equation} \hat{H}_{int}=e\frac{\partial H_{0}}{\partial\mathbf{p}}\cdot\mathbf{A}(\mathrm{\mathbf{x}},t)                                                                       \end{equation}

We assume a circularly polarized light:
\begin{equation}
\mathbf{E}(t)=E_c(\hat{x} \cos\omega t+ \hat{y}\sin \omega t)
\end{equation}
This corresponds to a vector potential
\begin{equation}
\begin{array}{c}
\mathbf{A}(\omega)=\frac{E_c}{\omega}(\hat{x}\sin\omega t-\hat{y}\cos\omega t)\\
=\frac{1}{2i}\frac{E_c}{\omega}((\hat{x}-i\hat{y})e^{i\omega t}-(\hat{x}+i\hat{y})e^{-i\omega t}),
\end{array}
\end{equation}

The current response to this field is:
\begin{equation}
j_{k}=\frac{e\tau}{\hbar\omega^2}E_{m}(\omega){\cal Q}_{kmn}(\mathbf{\omega})E_{n}^{*}(\omega).
\label{eq: total current}
\end{equation}
where $E_x(\pm\omega)=E_c/2,\,E_y(\omega)=\pm\frac{E_c}{2i}$. The ${\cal Q}_{kmn}(\omega)$ tensor, is given by integrating over the momentum resolved ${\cal Q}^{\alpha\beta}_{kmn}(\mathbf{k},\omega)$ as

\begin{equation}
{\cal Q}_{kmn}(\omega)=\int \frac{dk_xdk_y}{(2\pi)^2} \sum_{\alpha>0,\beta<0}{\cal Q}_{kmn}^{\alpha \beta}(\mathbf{k},\omega). \end{equation}

The momentum resolved ${\cal Q}^{\alpha\beta}_{kmn}(\mathbf{k},\omega)$ are in
turn given by

\begin{eqnarray}
{\cal Q}_{kmn}^{\alpha\beta}(\mathbf{k,\omega}) & = & \hat{\mathbf{x}_{k}}\cdot\left(\mathbf{v}_{\mathbf{k}}^{(\alpha)}-\mathbf{v_{k}}^{(\beta)}\right)M_{mn}^{\alpha\beta}(\mathbf{k})\nonumber\\  &  & \times2\pi \mathrm{\delta(E_{\mathbf{k}}^{(\alpha)}-E_{\mathbf{k}}^{(\beta)}-\hbar\omega)},
\end{eqnarray}

The matrix elements are given by

\begin{equation}
M_{mn}^{\alpha\beta}(\mathbf{k})=\left(e v_F\right)^2\langle\mathbf{k},\alpha|\sigma_{m}|\mathbf{k},\beta\rangle\langle\mathbf{k},\beta|\sigma_{n}^{\dagger}|\mathbf{k},\alpha\rangle\label{eq:traceformula}.
\end{equation}

In the following, we carry out the calculation for a non-monochromatic source
of light, which has an intensity ditribution as a function of angular
frequency, $I(\omega)$. The monochromatic limit is easy to extract, by
setting $I(\omega)$ to be proportional to a delta-function. The intensity of light at a given frequency with amplitude $E_x$ and $E_y$ is:
\begin{equation}
I(\omega)d\omega=\frac{1}{2}\epsilon_0 c (E_x^2+E_y^2)
\end{equation}
The $1/2$ comes from averaging the $\cos^2(\omega t),\,\sin^2(\omega t)$ over time. For the two circular polarizations of light this gives:
\begin{equation}
I(\omega)d\omega=\frac{1}{2}\epsilon_0 c (2E_{c+}^2+2E_{c-}^2)=\epsilon_0 c (E_{c+}^2+E_{c-}^2)=2\epsilon_0 c E_{c+}^2
\end{equation}
where we assumed that the two circular polarizations have the same
amplitude. So our circular polarization in terms of the solar
intensity is:
\begin{equation}
E_c^2=\frac{1}{2\epsilon_0 c}I(\omega)d\omega.
\end{equation}
Collecting all the coefficients, and using the
property of the tensor ${\cal Q}_{kmn}$, we get
\begin{equation}
j_{y}=\frac{e\tau}{{2}\hbar c \epsilon_0}\int d\omega {2}{\cal Q}_{yxy}(\mathbf{\omega})\frac{1}{4}\frac{I(\omega)}{\omega^2}.\label{eq: total current11}
\end{equation}


We now define $\eta(\omega)$ as a dimensionless quantity that encodes the
photocurrent respnse as a function of frequency, which also contains
all the intrinsic numerical factors:
\begin{equation}
\eta(\omega)=\frac{1}{2}\frac{1}{\left(e v_F\right)^2}\int\limits_{-q/2}^{q/2}\frac{dk_{x}}{2\pi
q}\int\limits_{-\infty}^{\infty}\frac{d\tilde{k}_{y}}{2\pi
q}\hbar v_F q\frac{{\cal Q}_{yxy}(\mathbf{k,\,\omega})}{v_F},
\end{equation}
Using this quantity in Eq.\ (\ref{eq: total current11}) , we get:
\begin{equation}
j_{y}=\frac{e^{3}\tau v_{F}^2q}{2c\epsilon_{0}\hbar^2}\int d\omega\eta(\omega)\frac{I(\omega)}{\omega^{2}}\label{eq: total current rescaled}
\end{equation}

Now let us substitute $I(\omega)$ for the sun. For a black body
at Temperature $T$, the black-body luminosity per $\omega$ is:
\begin{equation}
I(\omega,T)=\frac{1}{4\pi^2}\frac{\hbar \omega^3}{c^{2}}\frac{1}{\exp(\hbar\omega/k_{B}T)-1}
\end{equation}
where $I(\nu,T)$ is the energy per unit time (or the power) radiated
per unit area of emitting surface in the normal direction per unit
solid angle per unit frequency by a black body at temperature T.
The power per unit area arriving at the earth, and assuming normal incidence is:

\begin{equation}
I_{SE}(\omega,T_{sun})=\frac{R_{sun}^{2}}{R_{earth}^{2}}I(\omega,T_{sun})
\end{equation}

For low frequencies, we can approximate the black-body spectrum as
\begin{equation}
I(\omega,T)=\frac{k_{B}T\omega^{2}}{4 \pi^2 c^{2}}
\end{equation}
which is the Rayleigh-Jeans law.  Inserting this into Eq.~(\ref{eq: total current rescaled}), and taking
$\eta(\omega)=\eta$ (appropriate for large
frequencies), we get
\begin{equation}
j_{y}=\frac{e^{3}\tau
  v_{F}^{3}q^{2}}{2c\epsilon_{0}\hbar^2}\eta{\cal
  I}_{0}\frac{\omega_{max}}{v_F q}
\end{equation}
where the constant
\begin{equation}
{\cal I}_{0}=\frac{k_{B}T_{sun}}{c^{2}(2\pi)^2}\left(\frac{R_{sun}^{2}}{R_{earth}^{2}}\right)
\end{equation}

In order to extend the above analysis to finite temperature and chemical potential we must make an observation which ultimately lead to a simple modification of the formula above. For a system at finite temperature and with the chemical potential at an arbitrary point the incident light can excite or relax (through absorption or emission) electrons from any initial state to any final state. This must be accounted for in our model for the steady state $n_{\mathbf{k},\alpha}$. This physical considerations lead to a description identical to the one above, provided we use the following modified form for $\mathcal{Q}^{\alpha\beta}_{kmn}(\mathbf{k}, \omega)$
\beq
\mathcal{Q}^{\alpha\beta}_{kmn}(\mathbf{k}, \omega, T)=\mathcal{Q}^{\alpha\beta}_{kmn}(\mathbf{k}, \omega)(n_{\mathbf{k},\beta}^{0}-n_{\mathbf{k},\alpha}^{0})
\eeq



\begin{figure*}[!ht]
\centering
\begin{minipage}[b]{0.4\linewidth}
\includegraphics[width=7.5cm]{bandstructure_fig_new3_cropped.pdf}
\label{fig:minipage1f}
\includegraphics[width=8cm]{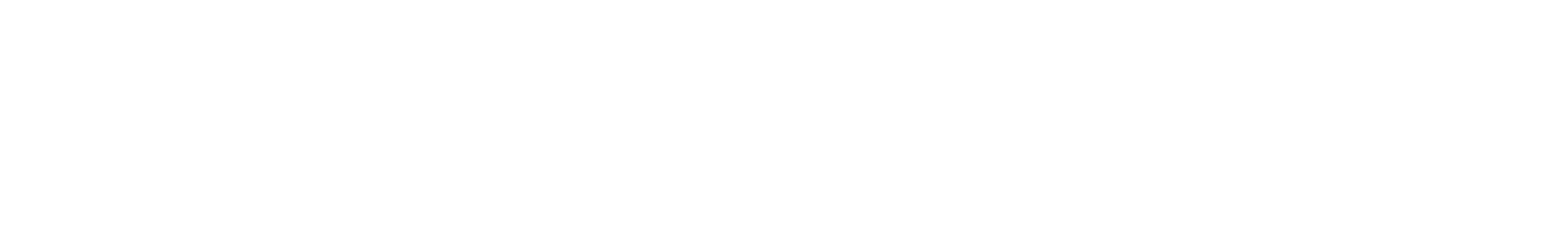}
\label{fig:minipage1p}\includegraphics[width=8cm]{blank_fig.pdf}
\label{fig:minipage1}
\end{minipage}
\quad
\begin{minipage}[b]{0.55\linewidth}
\includegraphics[width=11cm]{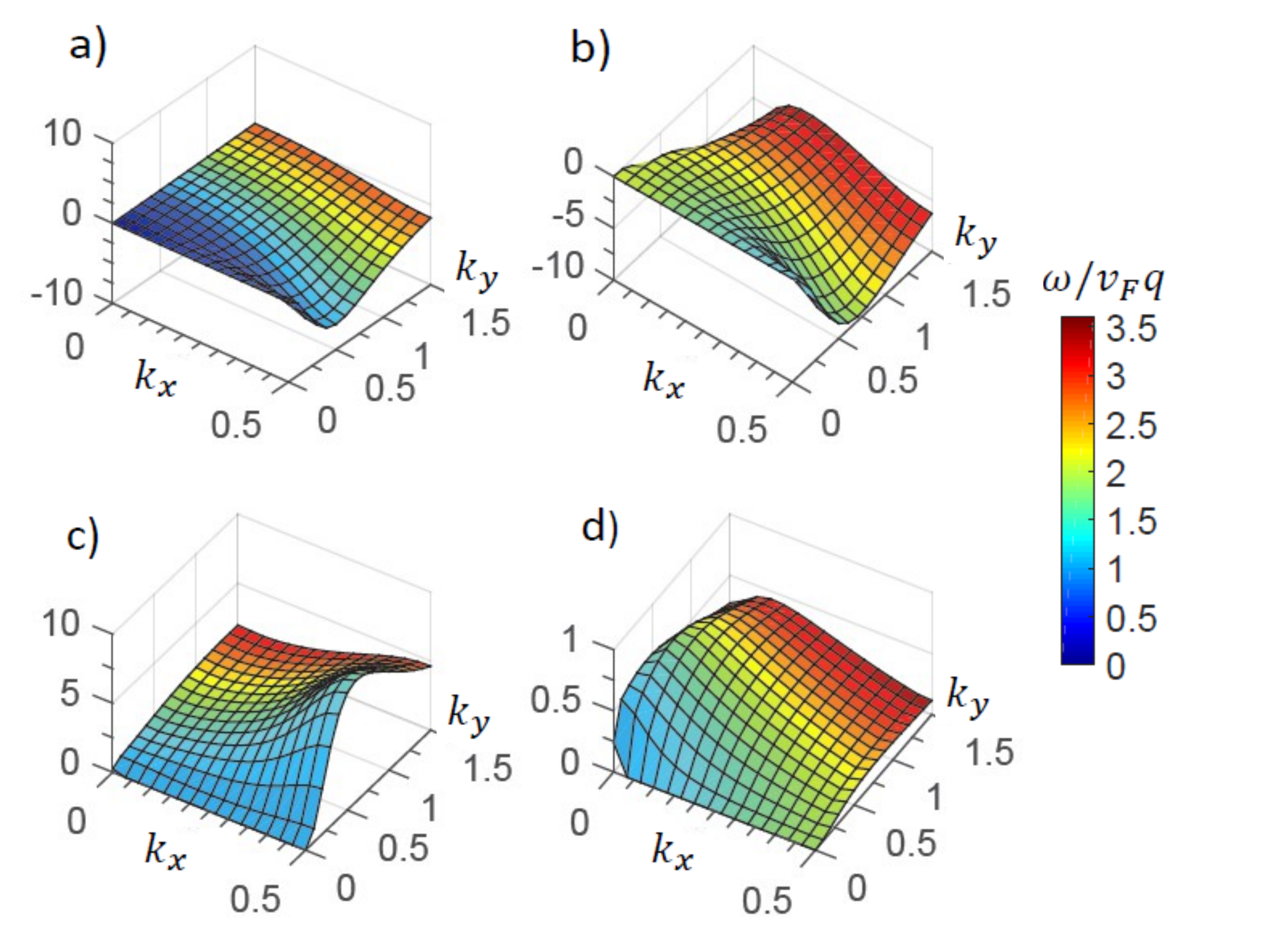}
\label{fig:minipage2}
\end{minipage}
\vspace{-0.8cm}
\caption{Top left: Effective bandstructure of the proposed heterostructure, cut along the line $k_{y}=0$ (units such that $|\mathbf{q}|=1$). Band indices are shown on the left. Transitions contributing to the summed momentum specific tensor ${\cal Q}_{kmn}^{\alpha\beta}(\mathbf{k})$ are depicted by arrows.
(a)-(d) Numerical results for $\int d\omega \tilde{{\cal Q}}_{yxy}^{\alpha\beta}(\mathbf{k})$ , in units of $\frac{e^{2}v_{F}^{3}}{c^{2}}$. The colors indicate the photon frequency of the transition as determined by the $\delta$ functions in Eq.\ (\ref{eq: q tensors 2nd order}).
Panels (a) and (b) contain the tensors for the transitions $(\alpha,\beta)=(1,-1)$ and $(2,-2)$,
respectively, in which the excitation leaves the electron momentum unchanged. These transitions indicated by green and purple
arrows, respectively, in the top left panel renormalize the conduction and valence bands.
(c) Same for $(\alpha,\beta)=(2,-1)$ and $(1,-2)$  (blue arrows in the top left panel).
(d) Same for $(\alpha,\beta)=(3,-1)$ and $(1,-3)$  (red arrows).}

\label{fig: Q tensors}.
\end{figure*}

\section{\MakeUppercase{Momentum specific response}}
In order to get some intuition for the origin of the photocurrent contributions, we study the momentum-specific response $\tilde{{\cal Q}}_{yxy}^{\alpha\beta}(\mathbf{k},\omega)$. This quantity allows us to understand which parts of the BZ contribute most to the effect.  This function is plotted in Fig.~\ref{fig: Q tensors} (a-d), where in addition to the response as a function of momentum, the photon energy responsible for the transition at each momentum is encoded in the color. We see that the effect is not exclusively due to the edges of the BZ. Rather, the contribution is uniformly distributed in momentum space, validating a perturbative perspective on the effects of the magnetic surface texture. Fig.~\ref{fig: Q tensors} also demonstrates that the sign change of $\eta(\omega)$ arises due to a sign difference between: (i) contributions of processes involving scattering by momentum $\pm\bq$ (indicated in red and blue in Fig.~\ref{fig: Q tensors}) which dominate for $\omega > v_F q$ and (ii) contributions of  momentum conserving processes (green and purple in Fig.~\ref{fig: Q tensors}) which dominate for $\omega< v_F q$

\section{\MakeUppercase{Perturbative analysis of the photocurrent response}}

In the following, we shall calculate the response tensor, accounting for the surface magnetic pattern within second order perturbation theory. This can be most conveniently expressed using momenta in the extended Brillouin zone. Denoting by $|\psi^{(0)}(\bk,\alpha)\9 $ the eigenstates of $H_0$ (without the magnetic structure), we expand the eigenstates in second order perturbation theory in $V=V^+ + V^-=(\textbf{u}\cdot\vec{\sigma})e^{i\bq\cdot r}+h.c.$, as
\begin{equation}
|\psi(\bk,\alpha)\9= |\psi^{(0)}(\bk,\alpha)\9 + |\psi^{(1)}(\bk,\alpha)\9+|\psi^{(2)}(\bk,\alpha)\9
\end{equation}
with $\alpha=c,v$ for conduction and valence bands. The first and second order corrections are given by
\begin{equation}
|\psi^{(1)}(\bk,\alpha)\rangle=\sum_{\sigma=\pm}\frac{1}{E^\alpha_\bk-H(\bk+\sigma\bq)}V^\sigma|\psi^{(0)}(\bk,\alpha)\rangle
\label{eq:psi1}
\end{equation}
and
\begin{eqnarray}
|\psi^{(2)}(\bk,v)\rangle&=&P^c_\bk\sum_{\sigma=\pm}\frac{1}{E^v_\bk-E^c_\bk}V^{\sigma\dagger}\frac{1}{E^v_\bk-H(\bk+\sigma\bq)}V^\sigma|\psi^{(0)}(\bk,v)\rangle.
\nonumber\\
|\psi^{(2)}(\bk,c)\rangle&=&P^v_\bk\sum_{\sigma=\pm}\frac{1}{E^c_\bk-E^v_\bk}V^{\sigma\dagger}\frac{1}{E^c_\bk-H(\bk+\sigma\bq)}V^\sigma|\psi^{(0)}(\bk,c)\rangle.
\label{eq:psi2} \nonumber\\
\end{eqnarray}
where $P_\bk^c$ is a projector on the conduction band state with momentum $\bk$. In second order perturbation theory, the total photocurrent response can be written as
\begin{equation}
{\cal Q}^{\rm ext}_{yxy}(\mathbf{k})  =   2\pi (e v_F)^2 \sum_{\lambda=0,+,-}\hat{\textbf{y}}\cdot\left(v^c_{\bk+\lambda\bq}-v^v_{\bk}\right) M_{xy}^{\lambda}(\mathbf{k}) \delta_\lambda(\omega,\mathbf{k})
\label{eq: q tensors 2nd orderp}
\end{equation}
where $v^v_\bk$ and $v^c_\bk$ denote the velocities in the conduction and valence bands, and the delta functions in Eq.~(\ref{eq: q tensors 2nd orderp}) were abbreviated to $\delta_\lambda(\omega,\mathbf{k})=\delta(E^c_{\mathbf{k}+\lambda \bq}+E^v_{\mathbf{k}}-\omega)$. The index $\lambda=0,+1,-1$ denotes process which correspond to  $ \bk,v \to \bk,c$, $ \bk,v \to \bk+\bq,c$ and $ \bk,v \to \bk-\bq,c$, respectively.
Our goal is to calculate the matrix elements:
\be
M_{mn}^{\lambda}(\mathbf{k})=\bra{\psi(\mathbf{k}+\lambda \bq,c)}\sigma_m|\psi(\bk,v)\9 \bra{\psi(\mathbf{k},v)}\sigma_n|\psi(\bk+\lambda\bq,c)\9,
\label{eq: M tensor 2nd order}
\ee
to second order in $V$. First, we describe transitions from $ \bk,v \to \bk+\bq,c$  in the extended BZ.  Two substitutions of $|\psi^{(1)}(\bk,v)\rangle$ from Eq.~(\ref{eq:psi1}) into Eq.~(\ref{eq: M tensor 2nd order}) yield

\beq
M^{+}_{mn}(\bk)=\sum_{r,s}u_r u^*_s\left[F^{v,v}_{mrsn}(\bk)+F^{v,c}_{mrns}(\bk)+F^{c,v}_{rmsn}(\bk)+F^{c,c}_{rmns}(\bk)\right]
\label{eq:Mp1m}
\eeq
where
\beq
F^{\alpha\beta}_{mnrs}(\bk)={\rm Tr}\left[P_{\bk+\bq}^c\sigma_{m}R_+^\alpha(\bk)\sigma_{n}P_\bk^v\sigma_{r}R_+^\beta(\bk)\sigma_{s}\right]
\label{eq: Ftensor}
\eeq
with $\alpha,\beta=v,c$ and
\begin{eqnarray}
R_\pm^v(\bk)&=&\frac{1}{E^v_\bk-H(\bk\pm\bq)}\nonumber\\
R_\pm^c(\bk)&=&\frac{1}{E^c_{\bk\pm\bq}-H(\bk)}
\end{eqnarray}
Note the permutation of the indices in Eq.~(\ref{eq:Mp1m}).


Next we compute the matrix elements for transitions which in the extended BZ, correspond to transitions $\bk \to \bk-\bq$. By taking $\bq \to -\bq$ in Eq.~(\ref{eq: Ftensor},
\beq
M_{mn}^{-}(\mathbf{k})=\sum_{r,s}u^*_r u_s \left[B^{v,v}_{mrsn}(\bk)+B^{v,c}_{mrns}(\bk)+B^{c,v}_{rmsn}(\bk)+B^{c,c}_{rmns}(\bk)\right]
\eeq
with
\beq
B^{\alpha\beta}_{mnrs}(\bk)={\rm Tr}\left[P_{\bk-\bq}^c\sigma_{m}R_-^\alpha(\bk)\sigma_{n}P_\bk^v\sigma_{r}R_-^\beta(\bk)\sigma_{s}\right].
\eeq


Next, we calculate the elements $M^{0}_{mn}$ which correspond to transitions $\bk,v \to \bk,c$. These can give a non zero contribution to the current in second order perturbation theory due to the renormalization of the bands, c.f. Eq.~(\ref{eq:psi2}). This yields

\begin{eqnarray}
M^0_{mn}(\bk)&=&\sum_{r,s}u_r^* u_s\Big\{\frac{1}{E_\bk^v-E_\bk^c}\Big(W^{+}_{nmrs}+W^{-}_{nmsr}+(W^{+}_{mnsr})^\dagger+(W^{-}_{mnrs})^\dagger\Big)\nonumber\\
&\phantom{=}&\phantom{blahblahbl}-\frac{1}{E_\bk^v-E_\bk^c}\Big(\tilde{W}^{+}_{mnrs}+\tilde{W}^{-}_{mnsr}+(\tilde{W}^{+}_{nmsr})^\dagger+(\tilde{W}^{-}_{nmrs})^\dagger\Big)\nonumber\\
&\phantom{=}&\phantom{blahblahbl}+Z^{+}_{mnrs}+Z^{-}_{mnsr}+(Z^{+}_{nmsr})^\dagger+(Z^{-}_{nmrs})^\dagger\Big\},
\label{eq: psi2 corrections}
\end{eqnarray}
where
\begin{eqnarray}
W^{\rho}_{mnrs}&=&{\rm Tr}\left[P_\bk^c\sigma_m P_\bk^c \sigma_n P_\bk^v \sigma_r R^v_\rho(\bk)\sigma_s\right],\nonumber\\
\tilde{W}^{\rho}_{mnrs}&=&{\rm Tr}\left[P_\bk^v\sigma_m P_\bk^v \sigma_n P_\bk^c \sigma_r \tilde{R}^c_{\rho}(\bk)\sigma_s\right],\nonumber\\
Z^{\rho}_{mnrs}&=&{\rm Tr}\left[P_\bk^c\sigma_m P_\bk^v \sigma_r R^v_\rho(\bk) \sigma_n \tilde{R}^c_{\rho}(\bk)\sigma_s\right],\nonumber\\
\end{eqnarray}
%

and where we have introduced the notation
\beq
\tilde{R}^c_\rho(\bk)=\frac{1}{E^c_{\bk}-H(\bk\pm \bq)}.
\eeq

In Eq.~(\ref{eq: psi2 corrections}), the first (second) term arises due to the second order corrections to the valence (conduction) states at momentum $\bk$, c.f. first (second) line in Eq.~(\ref{eq:psi2}). The third term in Eq.~(\ref{eq: psi2 corrections}) arises due to first order corrections (as in Eq.~(\ref{eq:psi1})) to both the valence and conduction bands.

To make a connection with the results presented in the main text, we would like to sum over momenta in the four quadrants of the BZ, and obtain the the momentum summed response tensor,
\begin{equation}
{\cal \widetilde{Q}}^{\rm ext}_{yxy}(\mathbf{k})  =  \sum_{\sigma,\sigma'=\pm}{\cal Q}^{\rm ext}_{yxy}(\sigma k_x,\sigma' k_y)
\label{eq: summed q tensors 2nd order}
\end{equation}

Note that the energy differences obey the symmetries
\begin{equation}
E^c_{(k_x,k_y)+\lambda \bq}-E^v_{(k_x,k_y)}=E^c_{(-k_x,k_y)-\lambda \bq}-E^v_{(-k_x,k_y)},
\label{eq: energy symmetries}
\end{equation}
and the velocities obey the symmetries appearing in Eq.~(17) of the main text. Using these symmetries, it is natural to define the functions $F_\lambda(\bk)$ which were used in Eq.~(20) the main text,
\begin{equation}
F_\lambda(\bk)=\sum_{\sigma,\sigma'=\pm}M^{\lambda\cdot\sigma}_{xy}(\sigma k_x,\sigma' k_y)\left(v_y^c(\bk+\lambda\bq)-v_y^v(\bk)\right)\sigma'
\end{equation}
The functions $F_\lambda(\bk)$ sum the matrix elements for the four transitions  $(k_x,\pm k_y) \to (k_x,\pm k_y)  +\lambda\bq$, and $(-k_x,\pm k_y) \to (-k_x,\pm k_y)  -\lambda\bq$. These transitions occur at the same photon frequency, by Eq.~(\ref{eq: energy symmetries}). Therefore, using the functions $F_\lambda(\bk)$, Eq.~(\ref{eq: q tensors 2nd orderp}) can be written as

\begin{equation}
{\cal \widetilde{Q}}^{\rm ext}_{yxy}(\mathbf{k})  =  2\pi {\textrm Re}\left\{{u_x u^*_z }\right\}(e v_F)^2 \sum_{\lambda=0,+,-} F_\lambda(\mathbf{k})\delta_\lambda(\omega,\mathbf{k})
\label{eq: q tensors 2nd order final result}
\end{equation}

\subsection{Second order perturbation theory in the reduced Brillouin zone scheme}
In this section, we will make the connetion between the response tensor ${\cal \widetilde{Q}}^{\rm ext}_{kmn}(\mathbf{k})$ obtained in second order perturbation theory,  and the response tensor ${\cal \widetilde{Q}}^{\rm \alpha \beta}_{kmn}(\mathbf{k})$ for the reduced Brillouin zone.
First, we note the relation between the \textit{unperturbed} eigenstates in the reduced BZ, which we denote by $|\bk,\alpha\9$, with $\alpha$ a positive (negative) integer for bands with $E>0$ ($E<0$), to those in the extended BZ, which we denote by $|\psi^{(0)}(\bk,a\9$, with $a=v,c$.  We will be interested only in the quadrant with $k_x,k_y>0$ due to the symmetries discussed in the main text.
\beq
\left|\bk,\alpha\right\9\to\left|\psi^{(0)}\left(\bk-(-1)^{\alpha}\lfloor\frac{\alpha}{2}\rfloor\bq,a\right)\right\9,\qquad
\label{eq: full to reduced BZ}
\eeq
where in the above equation, set $a=c$ for $\alpha>0$ and $a=v$ for $\alpha<0$.
\begin{table}[h]
\centering
 \begin{tabular}{|c|c|c|c|}
\hline
 \multicolumn{2}{|c|}{\phantom{blah}} &  \multicolumn{2}{|c|}{\phantom{blah}}   \\
 \multicolumn{2}{|c|}{Reduced zone  ${\cal \widetilde{Q}}^{\rm \alpha \beta}_{yxy}(\mathbf{k})$} &  \multicolumn{2}{|c|}{Extended zone ${\cal \widetilde{Q}}^\lambda(\bk_E)$}   \\
 \hline
$\alpha $ & $\beta$ & $\bk_E$ & $\lambda$ \\
\hline
$ 2n+3$ & $-(2n+1)$ & $\bk+n\bq$ &$+1$   \\
$2$ &$-1$ & $\bk $ & $-1$   \\
 $ 2n+1$ & $ -(2n+3) $ & $\bk+ (n+1) \bq  $ &$-1$  \\
 \hline
$ 2n+4$ & $ -2(n+2) $ & $-\bk+ (n+1) \bq $ &$+1$  \\
$ 1$ & $ -2 $ &$-\bk+\bq $ &$-1$   \\
$ 2n+2 $ & $ -(2n+4) $ & $-\bk+ (n+2) \bq$ &$-1$   \\
\hline
 $ 2n+1$ & $ -(2n+1) $ & $\bk+ n \bq  $ &$0$ \\
  $ 2n+2$ & $ -(2n+2) $ & $-\bk+ (n+1) \bq  $ &$0$  \\
  \hline
\end{tabular}
\caption{Mapping between the response tensors in the reduced Brillouin zone  ${\cal \widetilde{Q}}^{\rm \alpha \beta}_{yxy}(\mathbf{k})$ and the results obtained in second order perturbation theory. Only the values for the pairs $(\alpha,\beta)$ that have non zero rate in second order perturbation theory are shown. Note that the functions ${\cal \widetilde{Q}}^{\rm \alpha \beta}_{yxy}(\mathbf{k})$ are defined for momenta $\bk$ in the $k_x>0,k_y>0$ of the \textit{reduced} Brillouin zone. The value of these functions, in second order perturbation theory, corresponds to ${\cal \widetilde{Q}}^\lambda(\bk_E)$, where $\lambda$ and $k_E$ take the values shown in the table. In the left two columns, the $n$ is an integer such that $n\geq 0$.}
\label{table: map}
\end{table}
For the response second order perturbation theory, it is convenient to define each of the terms appearing in Eq.~(\ref{eq: q tensors 2nd order final result}) as
\begin{equation}
{\cal \widetilde{Q}}^{\lambda}(\mathbf{k})  =  2\pi {\textrm Re}\left\{{u_x u^*_z }\right\}(e v_F)^2 F_\lambda(\mathbf{k})\delta_\lambda(\omega,\mathbf{k})
\label{eq: q tensors 2nd order final result terms}
\end{equation}

From Eq.~(\ref{eq: full to reduced BZ}), we get a map between the response tensors ${\cal \widetilde{Q}}^{\rm \alpha \beta}_{kmn}(\mathbf{k})$ defined in the $k_x>0,k_y>0$ quadrant of the reduced Brillouin zone, to the processes corresponding to ${\cal \widetilde{Q}}^\lambda(\bk_E)$ in Eq.~(\ref{eq: q tensors 2nd order final result terms}), where $\bk_E$ takes value in the $k_x>0,k_y>0$ quadrant of the extended Brillouin zone. This map is constructed such that both $\bk_E$ and $\lambda$ are functions of $\bk$, $\alpha$ and $\beta$. This map is given explicitly in Table~\ref{table: map}.
\end{widetext}
\end{document}